\documentclass[pre,amsmath, twocolumn, showpacs, superscriptaddress]{revtex4-1}
\usepackage{graphicx}

\begin{document}

\title{Nonequilibrium Detailed Fluctuation Theorem for Repeated Discrete Feedback}

\author{Jordan M.\ Horowitz}
\affiliation{Departamento de F\'isica At\'omica, Molecular y Nuclear, Universidad Complutense de Madrid, 28040 Madrid, Spain}

\author{Suriyanarayanan Vaikuntanathan}
\affiliation{Institute for Physical Science and Technology, University of Maryland, College Park, MD 20742 USA}

\date{\today}

\begin{abstract}
We extend the framework of forward and reverse processes commonly utilized in the derivation and analysis of the nonequilibrium work relations to thermodynamic processes with repeated discrete feedback.
Within this framework, we derive a generalization of the detailed fluctuation theorem, which is modified by the addition of a term that quantifies the change in uncertainty about the microscopic state of the system upon making measurements of physical observables during feedback.
As an application, we extend two nonequilibrium work relations: the nonequilibrium work fluctuation theorem and the relative-entropy work relation.
\end{abstract}


\maketitle

\section{Introduction}

The \emph{nonequilibrium work relations} are a family of predictions concerning the fluctuations in the work performed on a microscopic system driven far from equilibrium \cite{Bochkov1977a,Jarzynski1997a, Crooks1998, *Crooks2000, Kawai2007,Jarzynski2006a,Jarzynski2007, Gomez-Marin2008a, Vaikuntanathan2009}.
They have been important for the study of fundamental issues in the thermodynamics of small systems and have proven to be powerful tools for calculating equilibrium free energy differences from nonequilbrium processes, both in experiments \cite{Collin2005,Liphardt2002,Douarche2005a} as well as in computer simulations~\cite{Chipot2007}.

At the heart of the nonequilibrium work relations is a statement about the time-reverseal symmetry of the microscopic dynamics termed the \emph{detailed fluctuation theorem} \cite{Gallavotti1995a,Evans2002a,Lebowitz1999,Hatano2001,Maes2003b,Seifert2005b,Harris2007} 
 (also called microscopic reversibility~\cite{Crooks2000} or the generalized fluctuation-dissipation theorem \cite{Bochkov1977a,Bochkov1981a,Astumian2007}).
 The detailed fluctuation theorem relates the probability to observe microscopic trajectories of the system through phase space during two thermodynamic processes related by time-reversal: the forward process and the reverse process.
This framework of forward and reverse processes has been beneficial for  investigating the role of irreversibility at the microscopic scale \cite{Jarzynski2007}.

However, the nonequilibrium work relations and the detailed fluctuation theorem are not applicable to systems manipulated using \emph{feedback} -- a procedure in which microscopic information about a system is utilized to manipulate or control its evolution.
Given the frequency with which feedback occurs in physics, biology, and engineering~\cite{Bechhoefer2005}, it is important to extend the  work fluctuation relations to include feedback.
This will clarify the thermodynamics of feedback \cite{Kim2004, Allahverdyan2008, Sagawa2008,Cao2009} as well as the the thermodynamics of computation \cite{Bennett1982, Piechocinska2000,Sagawa2009}, and possibly elucidate the role of  information processing in control theory~\cite{Cover, Touchette2000}.

Feedback can be implemented \emph{discretely} through a series of feedback loops initiated at a sequence of predetermined times or  \emph{continuously} at every instant of time.
Initial investigations into the work fluctuation relations in the presence of continuous feedback were made by Kim and Qian in the context of molecular refrigerators driven by velocity-depenedent feedback control \cite{Kim2007}.
The first work relation extended to include discrete feedback was the  nonequilibrium work fluctuation theorem~\cite{Jarzynski1997a}, recently reported by Sagawa and Ueda~\cite{Sagawa2010}.
They demonstrated that when a system is manipulated using one feedback loop the nonequilibrium work fluctuation theorem is modified by the addition of a term that accounts for  the microscopic information gained during feedback.
In this article, we develop a framework of forward and reverse processes for \emph{repeated} discrete feedback in order to analyze and extend Sagawa and Ueda's result.
Moreover, we generalize the detailed fluctuation theorem to include repeated discrete feedback.
We find that the information gained during feedback must be incorporated into the work relations.
As an application, we extend the nonequilibrium work fluctuation theorem \cite{Jarzynski1997a,Sagawa2010} as well as the relative-entropy work relation \cite{Kawai2007,Jarzynski2006a} in the presence of repeated feedback.
(While this article was under consideration, similar results were published~\cite{Ponmurugan2010}. We postpone a discussion comparing Ref.~\cite{Ponmurugan2010} with the present work until the conclusion.)

Our central result (Eq.~\ref{eq:dBal2} below) can be summarized as follows.
Consider a classical thermodynamic system initially in equilibrium at inverse temperature $\beta$.
Imagine driving this system away from equilibrium from time $t=0$ to $\tau$ by implementing a series of feedback loops at $N$ predetermined times $t_k$, $k=1,\dots, N$.
At each $t_k$, a physical observable $M_k$ is measured.
Based on the outcome of this measurement we drive the system by varying a set of external parameters $\lambda$ with time.
In each repetition or realization of this entire process, which we call the \emph{forward} process, the system will trace out a different microscopic trajectory $\gamma_{\tau,0}$ through phase space.
Furthermore, the protocol $\Lambda_t$ used to vary the external parameters $\lambda$ will differ in each realization due to fluctuations in the measurement process.
We are interested in comparing the statistics of $\gamma_{\tau,0}$ and $\Lambda_t$ in the forward process to those of the time-reversed conjugate pairs $\tilde\gamma_{\tau,0}$ and $\tilde\Lambda_t$ in the time-reversed process, which we call the \emph{reverse} process.
There is no feedback in  the reverse process (no measurements are made).
Instead an ensemble of realizations of the reverse process is generated by executing each external parameter protocol observed in the forward process in reverse.
Our main result is that the ratio of the probability to observe $\gamma_{\tau,0}$ and $\Lambda_t$ in the forward process $\mathcal{P}[\gamma_{\tau,0};\Lambda_t]$ to the probability to observe $\tilde\gamma_{\tau,0}$ and $\tilde\Lambda_t$ in the reverse process $\tilde{\mathcal P}[\tilde\gamma_{\tau,0};\tilde\Lambda_t]$ satisfies a \emph{detailed fluctuation theorem for discrete feedback}
\begin{equation}\label{eq:dBal2}
\frac{\mathcal{P}[\gamma_{\tau,0};\Lambda_t]}{\tilde{\mathcal P}[\tilde{\gamma}_{\tau,0};\tilde\Lambda_t]}=e^{\beta W_d[\gamma_{\tau,0};\Lambda_t]+I[\gamma_{\tau,0};\Lambda_t]},
\end{equation}
where $W_d$ is the dissipated work.
The new quantity appearing in Eq.\ \ref{eq:dBal2}, $I$, quantifies the change in our uncertainity about the microscopic state of the system upon measuring the physical observables $M_1, M_2, \dots , M_N$ in each realization.
The avereage of $I$ over many realizations $\langle I\rangle$ is the mutual information \cite{Cover}, which is an information theoretic measure of the reduction in our uncertainity about the microscopic state of the system upon making measurements.
Moreover, $\langle I\rangle$ naturally appears in the thermodynamics of feedback developed by Cao and Feito~\cite{Cao2009}, where $\langle I\rangle$ is equal to the reduction in the Shannon entropy of a thermodynamic system under feedback control [see Eq.~(7) of Ref.~\cite{Cao2009}].

Our analysis begins in Sec.~\ref{sec:motivation} by motivating the definitions of the forward and reverse process using the Szilard engine as a pedagogical example.
Our main result (Eq.~\ref{eq:dBal2}) is then derived in Sec.~\ref{sec:derivation}.
In Sec.~\ref{sec:I}, the interpretation of $I$ is developed in detail using ideas from Bayesian inference.
Equation \ref{eq:dBal2} is then exploited in Sec.~\ref{sec:applications} to generalize two work fluctuation relations, the nonequilibrium work fluctuation theorem \cite{Jarzynski1997a,Sagawa2010} and the relative-entropy work relation \cite{Kawai2007,Jarzynski2006a}.
Finally, we conclude in Sec.~\ref{sec:conclusion} with an outlook towards future research directions.

\section{Motivation and Definitions}\label{sec:motivation}

\begin{figure}[htb]
\includegraphics[scale=.25]{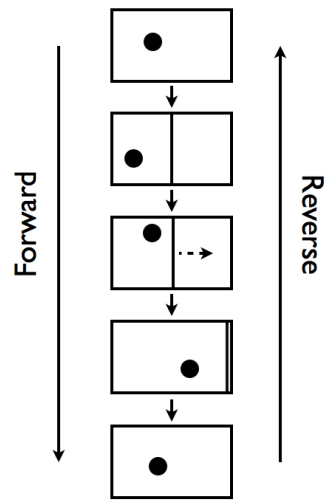}
\caption{Illustration of a realization of the forward and the reverse processes for the Szilard engine in which the particle is measured to be in the left half of the box.
The forward process is depicted by the sequence of illustrations running from top to bottom.  The reverse process is the time-reversed forward process; as such, time flows from bottom to top.}
\label{fig:szilard}
\end{figure}

Before deriving Eq.~\ref{eq:dBal2}, it is instructive to first motivate and establish the definitions of the forward and reverse processes in the context of the Szilard engine \cite{Szilard1964}, depicted in Fig.~\ref{fig:szilard}.
This will generalize the usual notions of forward and reverse processes common in the study of the work relations \cite{Crooks1998,Crooks2000,Kawai2007}.
The Szilard engine is composed of a single ideal gas particle in a box of volume $V$ in thermal contact with a heat bath at inverse temperature $\beta$.
We begin by describing the forward process, which is illustrated in Fig.~\ref{fig:szilard} by the sequence of snapshots proceeding from top to bottom.
Initially, the engine is allowed to relax to equilibrium.
A partition is then inserted in the center of the box, isolating the particle in either the left or right half of the box.
Feedback begins by measuring in which half of the box the particle is located.
The position of the partition is then shifted in a manner that depends on the measurement outcome: if the particle is found in the left (right) half of the box, the partition is slide all the way to the right (left).
Finally, the partition is removed and the particle is allowed to relax back to equilibrium.
Imagine repeating this process a number of times.
Each time equilibrating the particle, implementing the feedback loop, and finally allowing the engine to relax back to equilibrium.
This generates an ensemble of realizations of the forward process.

Within the framework of the work relations, the reverse process is implemented by carrying out each step (or each macroscopic control action) of the forward process in the reverse order.
For feedback processes the external parameter protocols are implemented in response to the outcomes of measurements.
The naive time-reversal of  this procedure -- implementing a protocol and then making the measurement used to determine this protocol -- would be acausal, because we would have to implement protocols in response to measurements made after the protocol was executed.
Instead, we generate an ensemble of realizations of the reverse process by first generating an ensemble of realizations of the forward process and then implementing the reverse of each protocol which was observed in the forward process. 
For example, suppose we observe a realization of the Szilard engine in which the particle was found in the left half of the box and the partition was moved to the right.
Having observed this realization of the forward process, we generate a realization of the reverse process by actuating each action of the forward process in reverse, which is depicted in Fig.~\ref{fig:szilard} by reading the images from bottom to top.
The particle is first equilibrated at inverse temperature $\beta$.
The partition is then inserted on the right side of the box and then slide to the center.
Finally, the partition is removed and the engine is allowed to relax back to equilibrium.
Repeating this process a number of times, each time reversing an observed realization of the forward process generates an ensemble of realizations of the reverse process.

Observe that in the reverse process no measurements are performed.
Instead, in an ensemble of realizations of the reverse process the protocols are implemented randomly according to the distribution in which they occur in the forward process.
The reverse process cannot be executed independently of the forward process; one must first perform the forward process.
This reliance of the reverse process on the forward process is a consequence of the time-reversal asymmetry of feedback and is an essential difference between thermodynamic processes with and without feedback.

\section{Derivation}\label{sec:derivation}

We are now in a position to derive Eq.~\ref{eq:dBal2}.
Let us begin by fixing notation.
Consider a classcial system, whose position in phase space (or microscopic configuration) is $z=({\bf x},{\bf p})$, where ${\bf x}$ denotes the system's coordinates and ${\bf p}$ denotes its momentum.
The energy of the system $E(z,\lambda)$ is parameterized by a vector of controllable external parameters $\lambda$ and is assumed to be time-reversal invariant for each fixed $\lambda$, $E(z,\lambda)=E(z^*,\lambda)$, where $z^*=({\bf x}, -{\bf p})$.
The dynamics are assumed to be Markovian dynamics (which includes deterministic dynamics) that preserve the canonical equilibrium distribution for each fixed $\lambda$
\begin{equation}\label{eq:peq}
P^{eq}(z|\lambda)=e^{\beta[F(\lambda)-E(z,\lambda)]},
\end{equation}
where $F(\lambda)$ is the free energy.
The position of the system at time $t$ will be denoted as  $z_t$.
The collection of phase space points visited by the system during the course of its evolution from $t=r$ to $s$ will be termed a \emph{microscopic trajectory} and will be labeled  $\gamma_{s,r}=\{z_t\}_{t=r}^s$.

The \emph{forward process} is defined as the following sequence of events.
The system is initially equilibrated with a thermal reservoir at inverse temperature $\beta$  with the external parameters fixed at $\lambda=A_0$.
Consequently, the initial statistical state of the system is  $P^{eq}(z|A_0)$~(Eq.~\ref{eq:peq}).
From $t=0$ to $t_1$ the system is driven away from equilibrium by varying  $\lambda$ with time using a predetermined initial protocol $\lambda_t^0$, from $\lambda_0^0=A_0$ to $\lambda_{t_1}^0=B_0$.
Then at subsequent times $t_k$, $k=1,\dots,N$, feedback loops are implemented.
At each $t_k$ a physical  observable $M_k$ is measured with (possibly continuous) outcomes $m_k$.
Each measurement outcome $m_k$ occurs with a probability that depends on the phase space position of the system at the time of measurement, $P_k(m_k|z_{t_k})$, and is independent of the previous measurements.
We collect all the measurement outcomes up to and including time $t_k$ into a vector $\mu_{k}=\{m_1,\dots,m_k\}$, which we call the the \emph{measurement trajectory} up to $t_k$.
During each time interval from  $t=t_k$ to $t_{k+1}$ ($t_{N+1}=\tau$) the external parameters are varied using a protocol which depends on the outcomes of all measurements up to $t_k$, $\lambda_t^k(\mu_k)$, from $\lambda_{t_k}^k(\mu_k)=A^k(\mu_k)$ to $\lambda_{t_{k+1}}^k(\mu_k)=B^k(\mu_k)$.
Additionally, we assume that each $\mu_k$ is associated to a unique protocol [\emph{i.e.}\ $\lambda_t^k(\mu_k)\neq\lambda^k_t(\mu^\prime_k)$ for all $\mu_k\neq\mu^\prime_k$] and that $A^k(\mu_{k})=B^{k-1}(\mu_{k-1})$, to ensure that the protocol is continuous at each measurement time $t_k$.
The microscopic trajectory $\gamma_{t_{k+1},t_k}$ taken by the system during this time  interval occurs with probability $P\left[\gamma_{t_{k+1},t_k}|z_k,\lambda_t^k(\mu_{k})\right]$, which is conditioned only on the position of the system at time $t_k$, $z_{t_k}$ -- since the dynamics are Markovian -- and depends on the protocol executed $\lambda^k_t(\mu_k)$.
The \emph{complete protocol} executed from $t=0$ to $\tau$, we represent by collecting the individual protocols used in each feedback loop into a vector, $\Lambda_t(\mu_N)=\{\lambda^0_t,\dots,\lambda^N_t(\mu_N)\}$.
The probability to observe a realization of the entire forward process with trajectory $\gamma_{\tau,0}$ and 
protocol $\Lambda_t(\mu_N)$ is 
\begin{widetext}
\begin{equation}\label{eq:forward}
\mathcal{P}\left[\gamma_{\tau,0};\Lambda_t\right]=
P\left[\gamma_{\tau,t_N}|z_{t_N},\lambda_t^N(\mu_N)\right]P_N(m_N|z_{t_N})\\
 \cdots P\left[\gamma_{t_2,t_1}|z_{t_1},\lambda_t^1(\mu_1)\right]P_1(m_1|z_{t_1})P\left[\gamma_{t_1,0}|z_0,\lambda_t^0\right]P^{eq}(z_0|A_0).
\end{equation}
\end{widetext}
The work done on the system along this trajectory is
\begin{equation}\label{eq:W}
\begin{split}
W[\gamma_{\tau,0};\Lambda_t]&=\sum_{k=0}^N W^k[\gamma_{t_{k+1},t_k};\lambda^k_t(\mu_k)] \\
&=\sum_{k=0}^N\int_{t_k}^{t_{k+1}}ds\, \dot\lambda^k_s(\mu_{k})
\frac{\partial}{\partial\lambda}E\left[z_s,\lambda^k_s(\mu_{k})\right],
\end{split}
\end{equation}
the heat flow into the system is 
\begin{equation}\label{eq:Q}
\begin{split}
Q[\gamma_{\tau,0};\Lambda_t]
&=\sum_{k=0}^N Q^k[\gamma_{t_{k+1},t_k};\lambda^k_t(\mu_k)] \\
&=\sum_{k=0}^N\int_{t_k}^{t_{k+1}}ds\, \dot{z}_s\frac{\partial}{\partial z}E\left[z_s,\lambda^k_s(\mu_{k})\right],
\end{split}
\end{equation}
and the change in energy satisfies the first law of thermodynamics
\begin{equation}\label{eq:delE}
\begin{split}
\Delta E[\gamma_{\tau,0};\Lambda_t]&=E[z_\tau,B^N(\mu_N)]-E(z_0,A_0) \\
&=W[\gamma_{\tau,0};\Lambda_t]+Q[\gamma_{\tau,0};\Lambda_t],
\end{split}
\end{equation}
where $t_0=0$ and $\lambda_t^0(\mu_0)=\lambda_t^0$. 
Since the protocols depend on the measurement outcomes the free energy difference is realization dependent, 
\begin{equation}
\begin{split}
\Delta F[\Lambda_t]&=F\left[\lambda^N_\tau(\mu_N)\right]-F\left[\lambda_0^0\right] \\
&=F[B^N(\mu_N)]-F(A_0).
\end{split}
\end{equation}
Likewise, the dissipated work is
\begin{equation}\label{eq:wDiss}
W_d[\gamma_{\tau,0};\Lambda_t]=W[\gamma_{\tau,0};\Lambda_t]-\Delta F[\Lambda_t].
\end{equation}

As discussed in Sec.~\ref{sec:motivation}, we generate an ensemble of realizations of the \emph{reverse} process by carrying out each observed realization of the forward process backwards in time.
Take for example the time-reversal of a realization of the forward process with protocol $\Lambda_t(\mu_N)=\{\lambda^0_t,\dots,\lambda^N_t(\mu_N)\}$.
The system is first equilibrated at inverse temperature $\beta$ with external parameters fixed at $\lambda^N_\tau(\mu_N)=B^N(\mu_N)$, so that the initial statistical state of the reverse process is $P^{eq}\left[z|B^N(\mu_N)\right]$ (Eq.~\ref{eq:peq}).
Then from time $t=0$ to $\tau$ the external parameters are varied according to the time-reversed individual protocols executed in the reverse order: for each time interval $t=\tau-t_{k+1}$ to $\tau-t_{k}$, $k=0,\dots,N$, the external parameters are varied according to the reverse individual protocol $\tilde\lambda^{N-k}_t(\mu_{k})=\lambda^k_{\tau-t}(\mu_k)$.
The \emph{reverse complete protocol} $\tilde\Lambda_t=\Lambda_{\tau-t}$ is $\tilde\Lambda_t(\mu_N)=\{\tilde\lambda_t^0(\mu_N),\dots,\tilde\lambda^N_t\}$.
Observe that in an ensemble of realizations of the reverse process the probability to observe reverse complete protocol $\tilde\Lambda_t(\mu_N)$, $\tilde\pi[\tilde\Lambda_t(\mu_N)]$, is independent of the microscopic trajectory and is equal to the probability that the conjugate forward complete protocol $\Lambda_t(\mu_N)=\tilde\Lambda_{\tau-t}(\mu_N)$ occurs in the forward process, $\pi[\Lambda_t(\mu_N)]$:
\begin{equation}\label{eq:probProt}
\tilde\pi[\tilde\Lambda_t]=\pi[\Lambda_t]=\int d\gamma_{\tau,0}\, \mathcal{P}[\gamma_{\tau,0};\Lambda_t],
\end{equation}
where $d\gamma_{\tau,0}$ is a measure on the space of microscopic trajectories.
Moreover, due to the assumed one-to-one correspondence between measurement trajectories and protocols, the probability distribution $\pi[\Lambda_t(\mu_N)]$ is equal to the probability distribution of measurement trajectories in the forward process
\begin{equation}\label{eq:probMeas}
\begin{aligned}
P_N(\mu_N)=P_{N}(m_N|\mu_{N-1})\cdots P_{2}(m_2|\mu_1)P_{1}(m_1),
\end{aligned}
\end{equation}
where $P_k(m_k|\mu_{k-1})$ is the conditional probability to observe measurement outcome $m_k$ in the forward process conditioned on the measurement trajectory $\mu_{k-1}$, and the equality follows from the product rule of conditional probabilities \cite{Caticha2008}.
Combining Eqs.~\ref{eq:probProt} and \ref{eq:probMeas}, the probability to implement reverse complete protocol $\tilde\Lambda_t$ in the reverse process is 
\begin{equation}\label{eq:probRevPro}
\tilde\pi[\tilde\Lambda_t(\mu_N)]=P_N(\mu_N).
\end{equation}

\begin{figure}[ht]
\includegraphics[scale=.25, angle=-90]{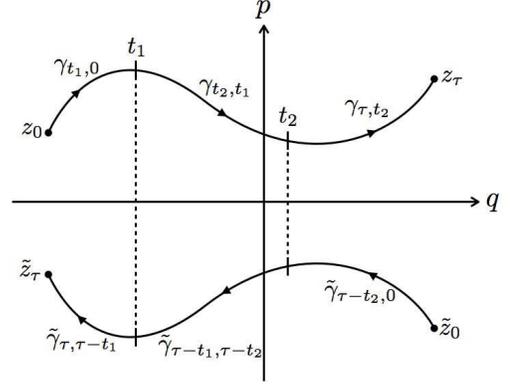}
\caption{Illustration of the forward trajectory $\gamma_{\tau,0}$ and reverse trajectory $\tilde\gamma_{\tau,0}$ with two feedback loops implemented at times $t_1$ and $t_2$.}
\label{fig:conjugate}
\end{figure}

For every trajectory from time $t=s$ to $r$, $\gamma_{r,s}=\{z_t\}_{t=s}^r$ there is a conjugate reverse trajectory $\tilde\gamma_{\tau-s,\tau-r}=\{\tilde{z}_t\}_{t=\tau-r}^{\tau-s}=\{z^*_t\}_{t=r}^s$, where $\tilde{z}_t=z^*_{\tau-t}$ (see Fig.~\ref{fig:conjugate}).
The probability to observe reverse trajectory $\tilde\gamma_{\tau,0}$ and reverse complete protocol $\tilde\Lambda_t$ in the reverse process is
\begin{eqnarray}
\label{eq:reverse3}
\mathcal{\tilde P}\left[\tilde\gamma_{\tau,0};\tilde\Lambda_t\right]
&=&P[\tilde\gamma_{\tau,0}|\tilde\Lambda_t]\tilde\pi[\tilde\Lambda_t],
\end{eqnarray}
where $P[\tilde\gamma_{\tau,0}|\tilde\Lambda_t]$ is the conditional probability to observe $\tilde\gamma_{\tau,0}$ conditioned on executing protocol $\tilde\Lambda_t$.
Substituting in Eqs.~\ref{eq:probMeas} and \ref{eq:probRevPro}, and expanding $P[\tilde\gamma_{\tau,0}|\tilde\Lambda_t]$ in conditional probabilities using the product rule of conditional probabilities \cite{Caticha2008}, allows us to express $\tilde{\mathcal P}$ as
\begin{widetext}
\begin{equation}
\label{eq:reverse}
\mathcal{\tilde P}\left[\tilde\gamma_{\tau,0};\tilde\Lambda_t\right]
=P[\tilde\gamma_{\tau,\tau-t_1}|\tilde{z}_{\tau-t_1},\tilde\lambda^N_t]P[\tilde\gamma_{\tau-t_1,\tau-t_2}|\tilde{z}_{\tau-t_2},\tilde\lambda^{N-1}_t(\mu_1)]P_{1}(\mu_1) \dots P[\tilde\gamma_{\tau-t_N,0}|\tilde{z}_0,\tilde\lambda^0_t(\mu_N)]P_{N}(m_N|\mu_{N-1})P^{eq}\left[\tilde{z}_0| B^N(\mu_N)\right].
\end{equation}
\end{widetext}

The structure of Eq.~\ref{eq:reverse3} (Eq.~\ref{eq:reverse}) suggests an alternative method for implementing the reverse process.
We randomly select a reverse protocol $\tilde\Lambda_t(\mu_N)$ according to the distribution $\tilde\pi[\tilde\Lambda_t(\mu_N)]$ (Eq.~\ref{eq:probRevPro}).
Next, we equilibrate the system at inverse temperature $\beta$ with external parameters fixed at $\tilde\Lambda_0(\mu_N)=B^N(\mu_N)$, drive the system away from equilibrium according to $\tilde\Lambda_t(\mu_N)$, and finally allow the system relax back to equilibrium at inverse temperature $\beta$ with external parameters fixed at $\tilde\Lambda_\tau(\mu_N)=A^0$.

With this setup, we can now derive Eq.~\ref{eq:dBal2} as a consequence of the \emph{detailed fluctuation theorem} \cite{Gallavotti1995a,Evans2002a,Lebowitz1999,Hatano2001,Maes2003b,Seifert2005b,Harris2007,Crooks2000,Jarzynski2006a,Bochkov1977a,Bochkov1981a,Astumian2007} 
\begin{equation}\label{eq:dBal}
\frac{P\left[\gamma_{t_{k+1},t_k}|z_{t_k},\lambda^k_t(\mu_k)\right]}{P\left[\tilde\gamma_{\tau-t_{k},\tau-t_{k+1}}|\tilde{z}_{\tau-t_{k+1}},\tilde\lambda^{N-k}_t(\mu_k)\right]}=e^{-\beta Q^k[\gamma_{t_{k+1},t_k};\lambda_t^k(\mu_k)]},
\end{equation}
where $Q^k$ is defined in Eq.~\ref{eq:Q}.
Equation~\ref{eq:dBal} has been derived for a wide class of dynamics and is a consequence of the time-reversal symmetry of the microscopic dynamics -- the energy is time-reversal invariant (see the discussion proceeding Eq.~\ref{eq:peq}).

\begin{figure}[htb]
\includegraphics[scale=.3, angle=-90]{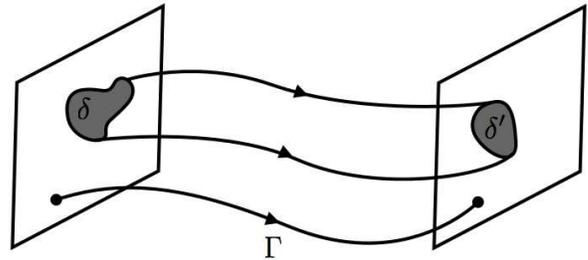}
\caption{Illustration of the tube of trajectories of the forward process evolving from phase space region $\delta$ and to region $\delta^\prime$ under  Hamiltonian dynamics driven by external parameter protocol $\lambda^\delta_t$ associated to measuring the initial state of the system inside region $\delta$.
$\tilde\Gamma$, the conjugate trajectory of $\Gamma$, exemplifies a trajectory of the reverse process whose conjugate trajectory $\Gamma$ cannot be realized in the forward process simultaneously with $\lambda^\delta_t$, since $\Gamma$ begins outside of phase space region $\delta$.}
\label{fig:examplepic}
\end{figure}

To derive Eq.~\ref{eq:dBal2}, we take the ratio of Eqs.~\ref{eq:forward} and \ref{eq:reverse}, then substitute in Eqs.~\ref{eq:peq}, \ref{eq:Q}, \ref{eq:delE}, \ref{eq:wDiss}, \ref{eq:dBal}, and the definition of the change of uncertainty
\begin{equation}\label{eq:I}
I[\gamma_{\tau,0};\Lambda_t]=\ln\left[\frac{P_N(m_N|z_{t_N})\dots P_2(m_2|z_{t_2})P_1(m_1|z_{t_1})}{P_{N}(m_N|\mu_{N-1})\dots P_{2}(m_2|\mu_1)P_{1}(m_1)}\right],
\end{equation}
which, after a short manipulation, leads to Eq.~\ref{eq:dBal2}, reprinted here for convenience,
\begin{equation}\label{eq:dBal3}
\frac{\mathcal{P}[\gamma_{\tau,0};\Lambda_t]}{\tilde{\mathcal P}[\tilde{\gamma}_{\tau,0};\tilde\Lambda_t]}=e^{\beta W_d[\gamma_{\tau,0};\Lambda_t]+I[\gamma_{\tau,0};\Lambda_t]}.
\end{equation}
Equation \ref{eq:dBal3} (Eq.~\ref{eq:dBal2}) is an extension of the detailed fluctuation theorem (Eq.~\ref{eq:dBal}) for systems driven away from equilibrium by repeated discrete feedback.
However, there is a fundamental difference between Eqs.~\ref{eq:dBal3} and \ref{eq:dBal} due to the inherent time-reversal asymmetry of feedback.
Since no measurements are made in the reverse process, there are microscopic trajectories and reverse complete protocols whose time-reversed conjugates do \emph{not} occur together in the forward process; that is, there exists a $\gamma_{\tau,0}$ with conjugate reverse trajectory $\tilde\gamma_{\tau,0}$, and $\Lambda_t=\tilde\Lambda_{\tau-t}$, such that $\mathcal{P}[\gamma_{\tau,0};\Lambda_t]=0$ and $\tilde{\mathcal P}[\tilde\gamma_{\tau,0};\tilde\Lambda_{t}]\neq 0$.
Consequently, the ratio $\mathcal{P}/\tilde{\mathcal P}$, appearing in Eq.~\ref{eq:dBal3}, is well-defined, but the reciprocal $\tilde{\mathcal P}/\mathcal{P}$ is not well-defined -- mathematically, we say $\mathcal{P}$ is absolutely continuous with respect to $\tilde{\mathcal P}$ ($\mathcal{P}\ll\tilde{\mathcal P}$) \cite{Koralov}, however the reverse is not true.
For example, consider a Hamiltonian system in which we implement feedback by making an error-free measurement at $t=0$ of whether the system is in a region of phase space $\delta$.
If the system is found in $\delta$, we drive the system with external parameter protocol $\lambda^\delta_t$.
The region of phase space $\delta$ then evolves deterministically to the region of phase space $\delta^\prime$, as illustrated in Fig.~\ref{fig:examplepic}.
In the reverse process the initial system state is sampled from a canonical distribution over all phase space.
Consequently, when the protocol $\tilde\lambda^\delta_t$ is executed, the system may evolve along a trajectory $\tilde\Gamma$ -- the conjugate trajectory of $\Gamma$ depicted in Fig.~\ref{fig:examplepic}, which begins outside of $\delta$ and terminates outside of $\delta^\prime$.
Clearly, the conjugate trajectory $\Gamma$ can never be observed in the forward process simultaneously with $\lambda^\delta_t$; $\mathcal{P}[\Gamma;\lambda^\delta_t]=0$ while $\tilde{\mathcal P}[\tilde\Gamma;\tilde\lambda^\delta_t]\neq0$.

\section{Interpretation of I}\label{sec:I}

We have mentioned that $I$ (Eq.~\ref{eq:I}) quantifies a change in uncertainty about the microscopic state of the system upon making measurements. 
We now provide an argument supporting that assertion using methods of Bayesian inference.
Our analysis begins by using Bayes's theorem \cite{Caticha2008} to rewrite Eq.~\ref{eq:I} in terms of the conditional probability distributions $\rho(z_{t_k}|\mu_k)$ to find the system at $z_{t_k}$ conditioned on the sequence of measurement outcomes in $\mu_k$,
\begin{equation}\label{eq:I2}
I[\gamma_{\tau,0};\Lambda_t(\mu_N)]=\sum_{k=1}^N\ln\left[\frac{\rho(z_{t_k}|\mu_k)}{\rho(z_{t_k}|\mu_{k-1})}\right].
\end{equation}
To interpret Eq.~\ref{eq:I2}, recall that probability distributions measure the degree of belief a \emph{rational} person has in the truth of a proposition, \emph{i.e.}\ they quantify our uncertainty~\cite{Caticha2008}.
For example, as rational statistical physicists, our uncertainty in the state of our system at time $t_1$, just prior to the first measurement, is 
\begin{equation}
\rho(z_{t_1})=\int^{z_{t_1}} d\gamma_{t_1,0}\, P\left[\gamma_{t_1,0}|z_0,\lambda_t^0\right]P^{eq}(z_0|A_0).
\end{equation}
Upon making a measurement, we gain information altering our beliefs and forcing us to update (or change) the probability distribution describing our state of knowledge about the system.
For example, suppose at time $t_1$ we measured $M_1$ and obtained outcome $m_1$.
We  have  gained some information and as rational beings we \emph{must} update our uncertainty $\rho(z_{t_1})$.
 Bayesian inference tells us that the new probability distribution describing our uncertainty -- the posterior probability distribution -- is  obtained from Bayes' theorem and is simply $\rho(z_{t_1}|m_1)$, the conditional probability for the system to be at $z_{t_1}$ given that the outcome of the measurement was $m_1$ \cite{Caticha2008}.
Comparing with Eq.~\ref{eq:I2}, we see that the $k=1$ term in the sum is $\ln\left[\rho(z_{t_1}|m_1)/\rho(z_{t_1})\right]$, the log of the ratio of the probability distributions before and after the measurement;  hence, it is a measure of how our uncertainty changes upon making a measurement.
Repeating this argument, we find that each term in the sum in Eq.~\ref{eq:I2} represents a change in our uncertainty upon making each new measurement.
Notice that $I$ can be positive or negative in any given realization, our uncertainty can decrease or increase.
However, the average of $I$ over many realizations $\langle I\rangle$ is always positive \cite{Cover}, reflecting that on average gaining information lowers our uncertainty.

\section{Applications}\label{sec:applications}

Equation \ref{eq:dBal2} immediately leads to two work relations (Eqs.~\ref{eq:expavg} and \ref{eq:relent} below).
It is a straightforward exercise using Eq.~\ref{eq:dBal2}  to show that 
\begin{equation}\label{eq:expavg}
\left\langle e^{-\beta W_d-I}\right\rangle=1,
\end{equation}
where the angle brackets denote an average of an ensemble of realizations of the forward process.
Equation~\ref{eq:expavg} is a generalization of the nonequilibrium work fluctuation relation of Sagawa and Ueda \cite{Sagawa2010} for multiple feedback loops. 
Similarly, Eq.~\ref{eq:dBal2} implies a generalization of  the relative-entropy work fluctuation relation \cite{Kawai2007,Jarzynski2006a}:
\begin{equation}\label{eq:relent}
D[\mathcal{P}||\tilde{\mathcal P}]=\beta\langle W_d\rangle +\langle I\rangle,
\end{equation}
where $D(f||g)=\int dx\, f(x)\ln[f(x)/g(x)]$ is the relative entropy, an information theoretic measure of the distinguishability of two probability distributions \cite{Cover}.

Furthermore, applying Jensen's inequality \cite{Cover} to Eq.~\ref{eq:expavg} or exploiting the positivity of the relative entropy~\cite{Cover} in Eq.~\ref{eq:relent}, one finds that
\begin{equation}
\beta\langle W_d\rangle+\langle I\rangle\ge 0,
\end{equation}
which can been viewed as a generalization of the second law of the thermodynamics in the presence of feedback \cite{Sagawa2010}.

\section{Conclusion}\label{sec:conclusion}

For systems driven by repeated discrete feedback, we have introduced a framework of forward and reverse processes.
We defined a reverse process in which  the steps of the forward process are carried out backwards in time.
As a consequence, we found that the change in uncertainty $I$ (Eq.~\ref{eq:I}) during each feedback loop must be incorporated when analyzing the thermodynamics of feedback.
$I$ is a natural generalization to repeated discrete feedback of the information measure utilized by Sagawa and Ueda in Ref.~\cite{Sagawa2010}.
Cao and Feito have also observed that the ensemble average $\langle I\rangle$ naturally occurs in their thermodynamics of feedback \cite{Cao2009}.
These observations support the conclusion that analyzing feedback using the framework of forward and reverse processes developed here may be beneficial to understanding the thermodynamics of feedback.

Exploiting Eq.~\ref{eq:dBal2}, we generalized the detailed fluctuation theorem (Eq.~\ref{eq:dBal2}), the nonequilibrium work fluctuation theorem (Eq.~\ref{eq:expavg}), and the relative-entropy work relation (Eq.~\ref{eq:relent}) to systems manipulated by repeated feedback.

The next step in understanding the thermodynamics of feedback is to incorporate feedback into the fluctuation relations \cite{Gallavotti1995a,Evans2002a,Lebowitz1999,Hatano2001,Maes2003b,Seifert2005b,Harris2007,Speck2005b,Kurchan1998,Crooks1999a} which are predictions about the fluctuations of thermodynamic quantities in far from equilibrium systems.
A first step in this regard has already been taken by Kim and Qian \cite{Kim2007}, who have analyzed the fluctuation relations in the presence of velocity-dependent feedback control.

While this article was under consideration, another paper proposing a detailed fluctuation theorem in the presence of feedback was published~\cite{Ponmurugan2010}.
Although the results are similar, our analysis contains a number of additional, important elements not found in Ref.~\cite{Ponmurugan2010}.
Our central result (Eq.~\ref{eq:dBal2}) applies to processes with multiple feedback loops, while Ref.~\cite{Ponmurugan2010} considers only a single loop.
We also provide a detailed physical interpretation of the reverse process, including a description of the procedure for executing that process, and a discussion of the asymmetry between the forward and reverse processes.
Finally, we give a physical interpretation for the change of uncertainty along a microscopic trajectory.

Moreover, we believe the main conclusions [Eqs. (11) and (13)] of Ref.~\cite{Ponmurugan2010} suffer from physical inconsistencies.
While Eq.~(11) of Ref.~\cite{Ponmurugan2010} assumes that feedback is implemented in both the forward and reverse processes, the protocol employed in the reverse process is acausal: it is executed in response to a measurement made in the future [cf.~Eq.~(8) of Ref.~\cite{Ponmurugan2010}].
Reference \cite{Ponmurugan2010} also investigates forward and reverse processes identical to those discussed in the present paper.
In this context Ref.~\cite{Ponmurugan2010} proposes a Crooks-type fluctuation relation for the joint distribution of dissipated work and change in uncertainty, $p(W_d,I)$.
This result is problematic: the change of uncertainty in the reverse process is ill-defined, since no measurements are made in the reverse process \footnote{private communication, anonymous referee}.

\acknowledgements
We are grateful to Chris Jarzynski for many stimulating discussions as well as J.~M.~R.~Parrondo for his helpful suggests.
We would also like to thank the anonymous referees for their insightful comments.
Jordan M.~Horowitz was supported by the American Recovery and Reinvestment Act (ARRA) funds through grant number ECCS 0925365 from the National Science Foundation and by Grant MOSAICO (Spain).  Suriyanarayanan Vaikuntanathan acknowledges support from the National Science Foundation (USA) under CHE-0841557.

%
\end{document}